# Octahedral dynamics and local symmetry in hybrid perovskite FAPbI$_3$ under thermal excitation

H. Joshi [a]*, K. C. Bhamu[b], A. Shankar [c], Rana Biswas [d] and M. Wlazło * [e]

[a] Department of Physics, SRM University Sikkim, Tadong, Gangtok, Sikkim, India-737102.

[b] Department of Physics, Mody University of Science and Technology, Lakshmangarh, Rajasthan, India-332311

[c] Condensed Matter Theory Research Lab, Department of Physics, Kurseong College, Darjeeling, India-734203.

[d] Electrical & Computer Engineering Department, Iowa State University, Ames, Iowa, 50011, USA.

[e] Chemical and Biological Systems Simulation Lab, Centre of New Technologies, University of Warsaw, Banacha 2C, 02-097 Warsaw, Poland.

*Corresponding author: himanshuabijoshi09@gmail.com ,

m.wlazlo@cent.uw.edu.pl

**Abstract**: Density Functional Theory (DFT) and ab initio molecular dynamics (AIMD) simulations have been employed to investigate the evolution of local motifs within the tetragonal phase of FAPbI$_3$ under thermal excitation. Our results reveal a distinct broadening in the distribution of PbI$_6$ octahedral volumes with increasing temperature, indicating a gradual breakdown of symmetry and emergence of diverse local environments. These octahedral volume distortions are primarily driven by the dynamic behaviour of the FA cation leading to softening of PbI$_6$ octahedra, evident from calculated mean octahedral volume and Pb-I-Pb bond angles. The examination of electronic structure confirmed that this dynamic structural phenomenon is directly responsible for the change in fundamental band gap value, highlighting the role of PbI$_6$ octahedra in modifying and modulating the electronic properties in FAPbI$_3$. The results demonstrate the microscopic origin of thermally induced dynamical behaviour to the macroscopic electronic properties and underscore the pivotal role of local motifs in hybrid perovskites.

**Introduction:**

Organic-inorganic hybrid halide perovskite of the general formula $APbX_3$ (A = $CH_3NH_3$ or $CH(NH_2)_2$ and X = halide atom) have recently drawn widespread attention due to its superior solar power conversion efficiency [1-3]. These perovskites outperform silicon and other conventional photovoltaic materials [4, 5], making them particularly desirable for high performance photovoltaic applications. Significant advancements have demonstrated that substituting methylammonium (MA) = $CH_3NH_3$ by a large formamidinium (FA) = $CH(NH_2)_2$ cation, leads to remarkable power conversion efficiency above 22 % [6]. This improvement results because FA increases the effective cation radius, leading to a reduction in the optical band gaps and thereby enhancing light absorption [7-9]. In addition to this electronic influence, the FA cation plays a structural role by templating a network of corner-sharing $PbX_6$ octahedra. Thus, the orientation and dynamic behaviour of this molecular cation critically affect both stability and efficiency. Moreover, hybrid halide perovskites like $FAPbI_3$ are known to exhibit complex phase transitions, which is highly sensitive to temperature and pressure [10]. At ambient temperature, the compound stabilizes in a pseudo-cubic or α-phase [11], which transitions to tetragonal and subsequently to orthorhombic phase as temperature decreases [9, 12]. These lower temperature phases have a large band gap [13], making them poor solar absorbers and are therefore less explored in photovoltaic studies. However, the precise nature of phase transition and magnetic ordering in the low temperature phases of $FAPbI_3$ remains debatable [14-17]. Experimental evidences suggest that the orientation of the molecular cation drives phase transition in the hybrid halide perovskites [15], although a detailed theoretical understanding of the organic molecule orientation dynamics in these materials is still lacking. Interestingly, some studies have addressed the inconsistencies between theoretical predictions and experimental findings by employing a symmetry-broken polymorphous network approach [18, 19], particularly in band gap estimations and mixing enthalpies of cubic phase as well as other phases. Such networks, obtained by removing the standard restriction to a minimal unit cell size, allow for a variety of octahedral tilts and distortions, with significant lowering of calculated total energies [20, 21]. Although key aspects of polymorphous models for certain quantum materials have been established, the influence of temperature on local structural motifs and their effect on the electronic properties is not well understood and remains largely under-explored. Motivated by this, the present study investigates the role of temperature in driving octahedral distortions within the tetragonal phase of $FAPbI_3$. A key factor is the complexity in the behaviour of the A-site cation. Unlike in purely inorganic halide perovskites, where A-site is occupied by a spherical inorganic ion, the presence of an organic molecule introduces additional degrees of freedom. The FA cation, in particular, undergoes dynamic motions including dihedral rotations and high-frequency vibrations [22], which considerably influence both the local structural symmetry and properties of these materials.

In this work, we investigate the dynamical distortion in $PbI_6$ octahedral volumes and the Pb-I-Pb bond angles, induced due to FA molecular cation as a function of temperature. The electronic band structures were calculated at various temperature to correlate the effect of $PbI_6$ octahedral distortion with the change in electronic properties. To obtain a reasonable electronic structure of the large supercells used in our calculations, the band unfolding technique [23-26] was implemented, which maps the electronic states of the supercell Brillouin zone on to that of the primitive cell. The 0 K ideal-symmetry structure exhibits a lower band gap values than the temperature induced symmetry-broken structures. We show that thermal disorder first breaks the symmetry, which increases the gaps, before the effect of octahedral distortion and expansions take over. The FA cation orientation was examined up to a temperature range of 500K, revealing that the cation rotates more freely with increasing temperature, providing direct evidence of the observed broader octahedral distortions resulting from breaking of the local symmetry. Ab initio molecular dynamics (AIMD) further demonstrate that the local motifs evolve continuously with temperature leading to emergence of polymorphous network of octahedral tilts and distortions. The temperature driven evolution of local structural motifs and their influence on the electronic properties of hybrid halide perovskite materials remains an area that have not been thoroughly investigated. These insights linking the microscopic origins of dynamical behaviour to the macroscopic electronic properties in FA-based perovskites are crucial for the rational design of more efficient materials.

**Computational Methods:**

All calculations are based on first principles method and were performed employing the Vienna Ab Initio Simulation Package (VASP) [27, 28], incorporating projector augmented-wave (PAW) pseudopotentials [29] and employing the SCAN [30] exchange–correlation functional. The SCAN functional has shown improved accuracy in capturing structural and electronic properties especially in systems involving symmetry breaking [20], when compared to more conventional functionals such as PBE [31] or even DFT + U [32] approaches. The energy cutoff values were fixed to 500 eV for structural relaxation and increased to 550 eV for volume optimization and molecular dynamics (MD) simulations. Initial structural optimization was accomplished using a k-point density of approximately 1000 points per reciprocal atom, which was refined to 10,000 for the final geometry optimization. For AIMD simulations, calculations were carried out using only the Γ-point. Structural relaxation was set on a 2×2×2 supercell of the tetragonal symmetry with 384 atoms, until all atomic forces fell below 0.01 eV/Å. To explore the structural effect with temperature, AIMD simulations were carried out under constant number of atoms (N), pressure (P), and temperature (T) conditions using the NPT ensemble. Temperature and pressure control were maintained using the Langevin thermostat in conjunction with the Parrinello–Rahman barostat [33, 34]. A timestep of 1 fs was used throughout. The atomic degrees of freedom were subjected to a friction coefficient of 3 ps$^{-1}$, while the lattice degrees of freedom were governed by a coefficient of 10 ps$^{-1}$. Post-processing and visualization of structural and electronic data were performed using the Python Materials Genomics (pymatgen) [35] toolkit and the VESTA [36] software suite.

**Results and Discussion:**

The tetragonal phase of $FAPbI_3$ features corner sharing octahedra, where Pb atoms occupy the B-site and are coordinated by six iodine neighbour atoms. These octahedral are a defining feature of the perovskite structure. We consider the conventional 4 f.u. /cell minimal tetragonal structure [19, 37] and relax the atomic positions maintaining the symmetry. As previously discussed, to capture full range of the possible octahedral distortions, including Glazer type rotational modes like $a^0a^0b^-$, structural relaxation on a 2 × 2 × 2 supercell containing 32 f.u./cell was considered, which is sufficient for effective inclusion of long-range interactions [18, 38]. The optimized tetragonal 32 f.u. /cell structure showed limited FA molecular orientation only along the [100] or the [101] lattice direction and modest $PbI_6$ tilting (Figure 1a), implying that the symmetry is not broken. We use this structural configuration to enables us monitor the progressive weakening of structural disproportionation, offering insight into the underlying mechanisms of structural disproportionation and shedding light on how temperature influences the arrangement of local motifs. With thermal excitation introduced by AIMD, the displacement in the Pb atomic sites induces tilting of the $PbI_6$ octahedra, while the rotation of the FA molecule within its cage causes disproportionation among the octahedra (Figure 1b). Such symmetry breaking introduced by temperature severely affects the band gap of the material. It also causes the loss of structural bond which changes the local environment in the crystal symmetry leading to substantial energy lowering, thus stabilizing the structure.

To investigate how temperature influences the arrangement of local structural motifs, ab initio molecular dynamics (AIMD) simulations were carried out within the NPT ensemble. Although related works have been conducted across various material systems [39-41], essential simulation parameters are often not clearly specified, particularly local motif distributions are based on very short AIMD runs, sometimes only a few picoseconds. We focus primarily on structural motifs, which can be effectively examined through experimental methods such as synchrotron x-ray diffraction (s-XRD) [42] or pair distribution function analysis via local probes [43]. Some studies report the ferromagnetic nature of low temperature phases in $FAPbI_3$ [44, 45] and to investigate such behaviour, spin motif distributions have to be considered. However, such investigations require more sophisticated theoretical treatment to capture the behaviour of multiple magnetic sublattices. Therefore, spin motif distributions are not considered in the present discussion.

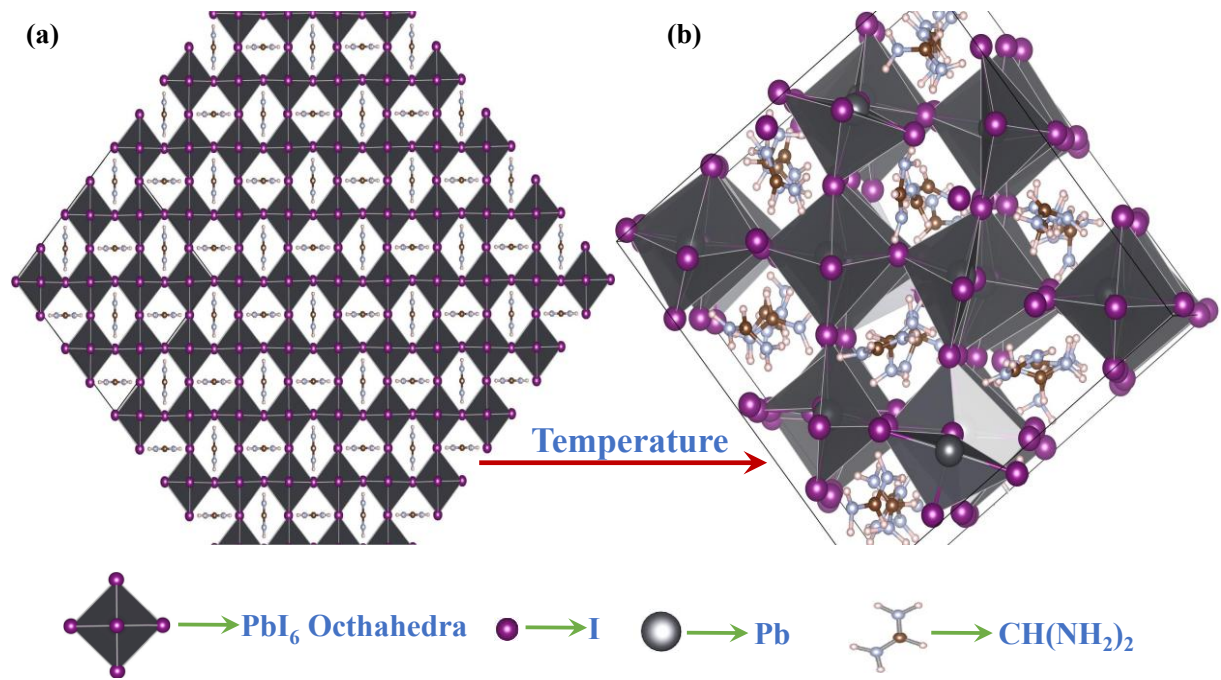

**Figure 1:** Effect of temperature on the FA molecular orientation and octahedral disproportionation. **(a)** Optimized tetragonal symmetry structure at 0 K with no octahedral tilting and minimal FA orientation. In 0 K, the cuboctahedra cages are extended along either the [100] or [010] orientation. **(b)** Snapshot of the structural evolution with temperature at 500 K. The crystal structure shows enhanced FA molecular rotation and octahedral tilting with the cuboctahedra cages at random orientation.

In AIMD simulations, atomic velocities are typically initialized based on a random distribution corresponding to the target temperature. This results in a system that begins in a non-equilibrium state, not yet fully adapted to external constraints like temperature or pressure, nor to internal atomic or molecular interactions. As a result, a finite equilibration time is necessary [46]. Conventionally, equilibrium is assessed by tracking the evolution of quantities such as lattice constants over time. Such plots of parameter evolution are crucial in understanding the physical state of the system with increasing temperature. Minimal fluctuations in the cell parameter even at higher ranges of temperature, as shown in figure 2, indicate that the system does not attain unphysical crystal structure due to lattice expansion as temperature rises.

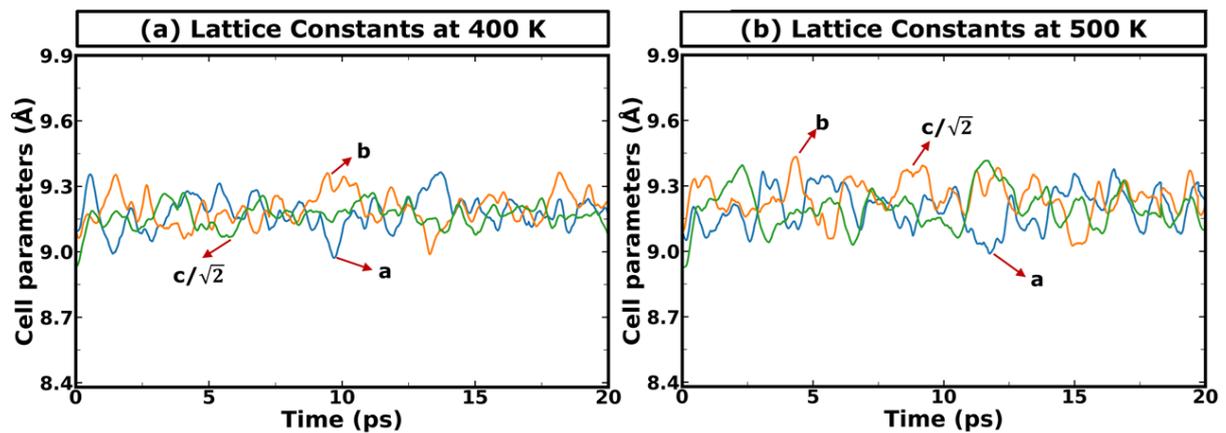

**Figure 2:** Ab initio molecular dynamics simulation of time evolution of lattice constant variations at higher ranges of temperature. The lattice constants evolution is calibrated up to 25 ps time, with the initial 5 ps reserved for system equilibrium and is hence omitted in the figure.

In the absence of thermal influence, the structure exhibits a uniform local environment, with negligible variation in the local structural motifs, as evident in Figure 3. At this point, the $PbI_6$ octahedra maintain a consistent volume, indicating a single, well-defined contribution to the structural framework. However, as temperature is introduced into the system, we begin to observe a noticeable distribution in these local motifs, quantified here through the calculated volumes of the $PbI_6$ octahedra. At elevated temperatures, the previously singular peak in the volume distribution broadens and shifts, suggesting that thermal fluctuations promote local distortions within the structure. This shift gradually evolves into a broad, Gaussian-like profile, marking the emergence of diverse local environments, as evident from the mean and standard deviation value of the octahedra volume obtained at different temperature. Mean value of the volume broadens with temperature, indicating softening of the octahedra due to tilting and stretching. For instance, the distribution observed at 500 K is significantly broader than that at 400 K, which in turn is broader than at 300 K, clearly revealing a temperature-dependent expansion of local motif diversity. At 0 K, the observation of a single local environment likely stems from the fact that the initial structure was not symmetry broken. We anticipate that if symmetry breaking were induced even at 0 K, by applying a small perturbation to slightly shift atoms from their equilibrium lattice sites (polymorphous structure), the structure would have reveal at least a dual local environment, attributable to distortions in the $PbI_6$ octahedra arising from the varied orientations of the FA cations confined within their cages [47]. The undistorted octahedra volume calculated at 0 K was 43.57 Å, which increases to a mean value of 46.58 Å at 500K. This is the structural figure print of increased dynamical disorder that represents local symmetry breaking. To the best of our knowledge, no direct theoretical or experimental comparison on PbI$_6$ octahedra volumes were obtainable from the literature, herein, we present the first comprehensive analysis.

Understanding the nature of these distributions, especially how they evolve with temperature, can offer key insights into the mechanisms driving disproportionation in hybrid perovskites. The broadened distribution at higher temperatures suggests a dynamic restructuring of local environments, potentially underpinning the material's phase evolution. While AIMD simulations offer a useful approximation of these behaviours, the absolute volume values should be interpreted cautiously; what holds greater significance is the pattern and extent of the distribution itself. Furthermore, capturing such subtle symmetry-breaking features experimentally would require advanced local probes, as conventional X-ray diffraction techniques may not be sensitive enough to detect them. This highlights the importance of complementary methods such as pair distribution function (PDF) analysis, which can more effectively reveal hidden local structural variations and provide a more comprehensive understanding of the material's thermally driven transformations.

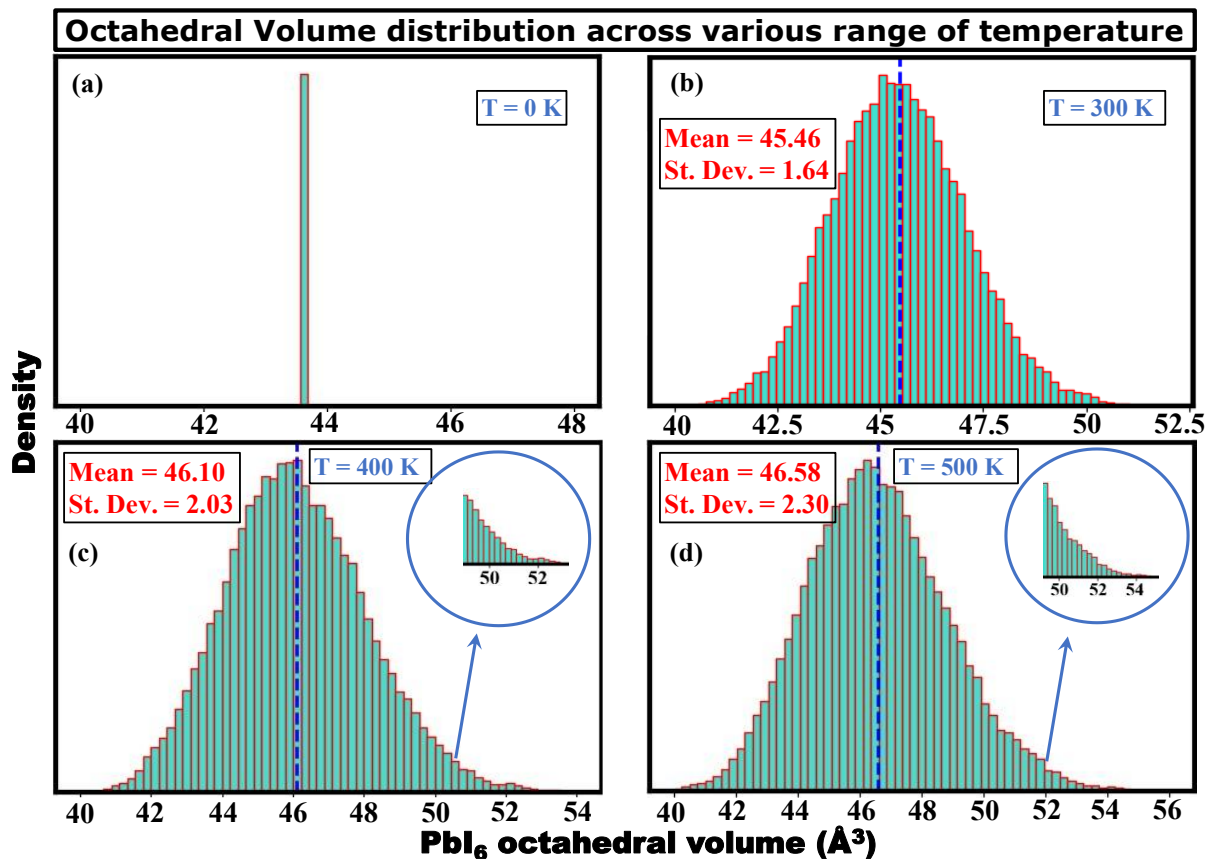

**Figure 3:** PbI$_6$ octahedral volume distributions at various temperatures, sampled over 20 ps following an initial 5 ps equilibration period. **(a)** Single local environment denoted by a single volume of the PbI$_6$ octahedra. **(b)**, **(c)** and **(d)** shows the broadening of the octahedral volume with temperature. The broadening indicates change in local crystal environment with temperature. The figure inserts on **(c)** and **(d)** highlight the enhanced volume distribution induced by temperature.

To understand the evolution of band structure with this dynamical distortion, unfolded band structure plotted in the first Briollouin zone across various range of temperature is shown in figure 4. The effective elemental contribution as well as the direct nature of the gap remains unchanged with the introduction of temperature. However, a non-monotonic evolution of band gap is observed, $E_g$ (0 K) < $E_g$ (500 K) < $E_g$ (400 K) < $E_g$ (300 K), highlighting the profound impact of octahedral distortion on the electronic landscape. At 0 K, the high symmetry structure yields well defined unfolded electronic states with a direct gap of 1.53 eV along the Γ-symmetry point, resulting in clean bands and aligns with the available literature [48].

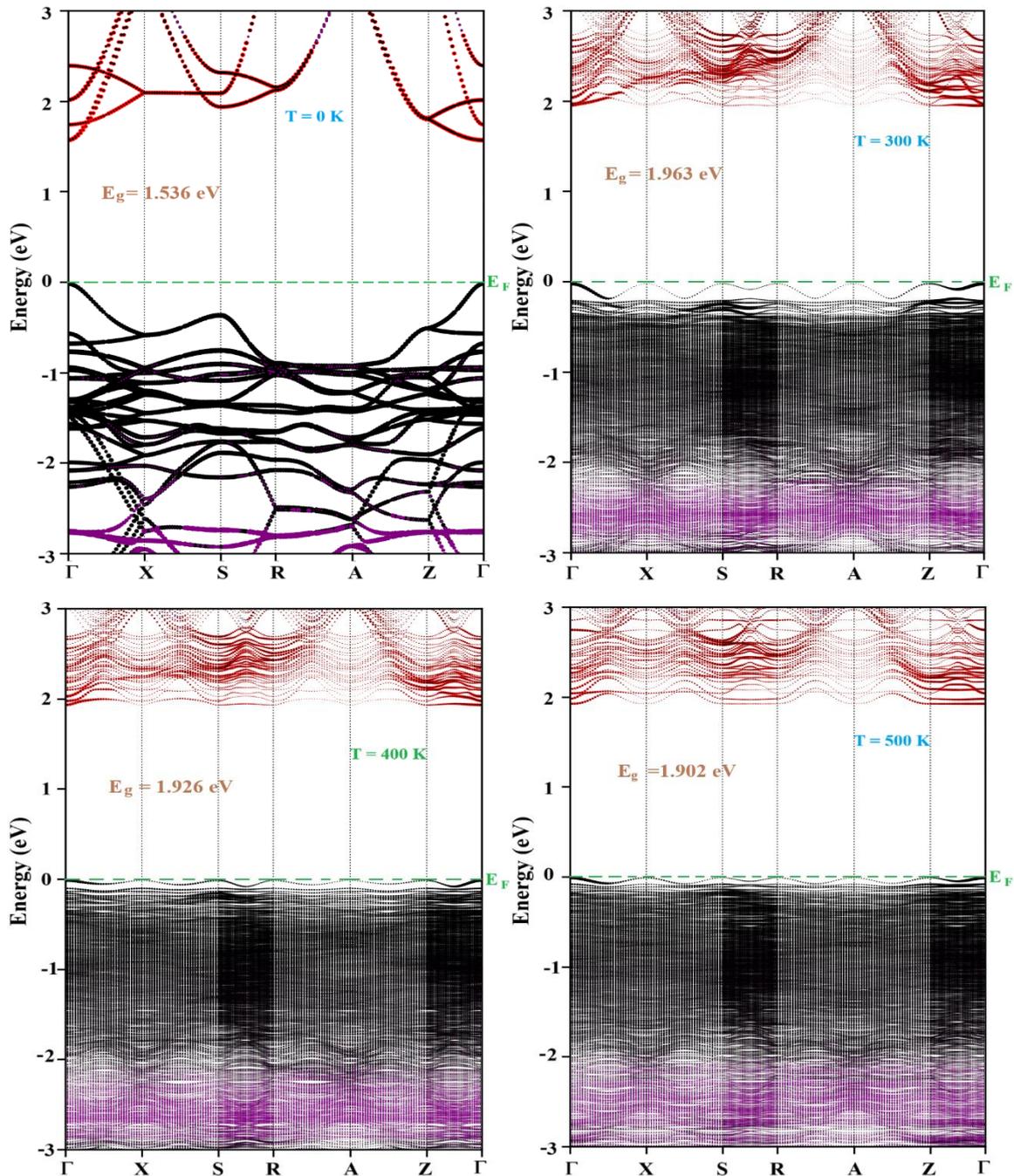

**Figure 4:** Unfolded effective band structure of 2 × 2 × 2 supercell containing 32 f.u./cell at 0K, 300 K, 400 K and 500 K temperature. Colour variation red and black in the band structure denote the elemental contribution to the band edges by Pb and I atoms respectively.

As temperature increases a significant broadening of spectral weights, particularly in the valence band region emerge. The highly smeared band edges provide a direct signature of the loss in symmetry induced by octahedral volume distortion and anharmonic FA cation motion. The FA molecule reorients freely, broadening the volume distribution as temperature increases resulting in significant narrowing of the band gap with increasing dynamic disorder. This trend reflects the competition between symmetry breaking, which promotes gap widening and disorder induced distortions, which counteract by reducing the gap. Also, a significant decrease in the Pb-I-Pb bond angle is observed as a result of dynamic disorder, shown in figure 5. These angles define the volume of the cuboctahedral cavity, where the ideal value of 180° corresponds to the most ordered cubic symmetry characterized with collinear bonds in Pb-I-Pb [49]. In absence of dynamical disorder, higher values of the bond angle are associated with the lower values in the band gap, as observed from our 0 K results. However, as temperature influences, the Pb-I-Pb bond angles become more broader due to enhanced rotation of the FA molecule. This additional rotational freedom induces further distortion to the $PbI_6$ octahedra, which in turn enhances FA molecular tilt angles and generates dynamic disorder, characterized by broadening of octahedral volumes. Thereby, the narrowing of band gap is observed between 300-500 K temperature. Interestingly, the band gap at 300K is larger in comparison to 0 K. This is primarily due to enhanced symmetry breaking induced by dynamic effect and is consistent with earlier reports [50]. The associated spectral blue shift at 300 K occurs because of the $PbI_6$ octahedral volume distribution from a sharp narrow peak at 0 K to a broadened distribution at 300 K (Figure 3). The significant non-rigid breathing-mode distortion, breaks the local symmetry removing electronic degeneracies that widens the band gap. The orbital overlap between the Pb and I electronic state, resulting from symmetry breaking, reduces the anti-bonding character at the valence band maxima thereby contributing to the observed electronic behaviour. However, as temperature increases, the lattice expansion dominates over the dynamic symmetry breaking, typically effecting the conduction band by lowering its energy, resulting in reduction of band gap.

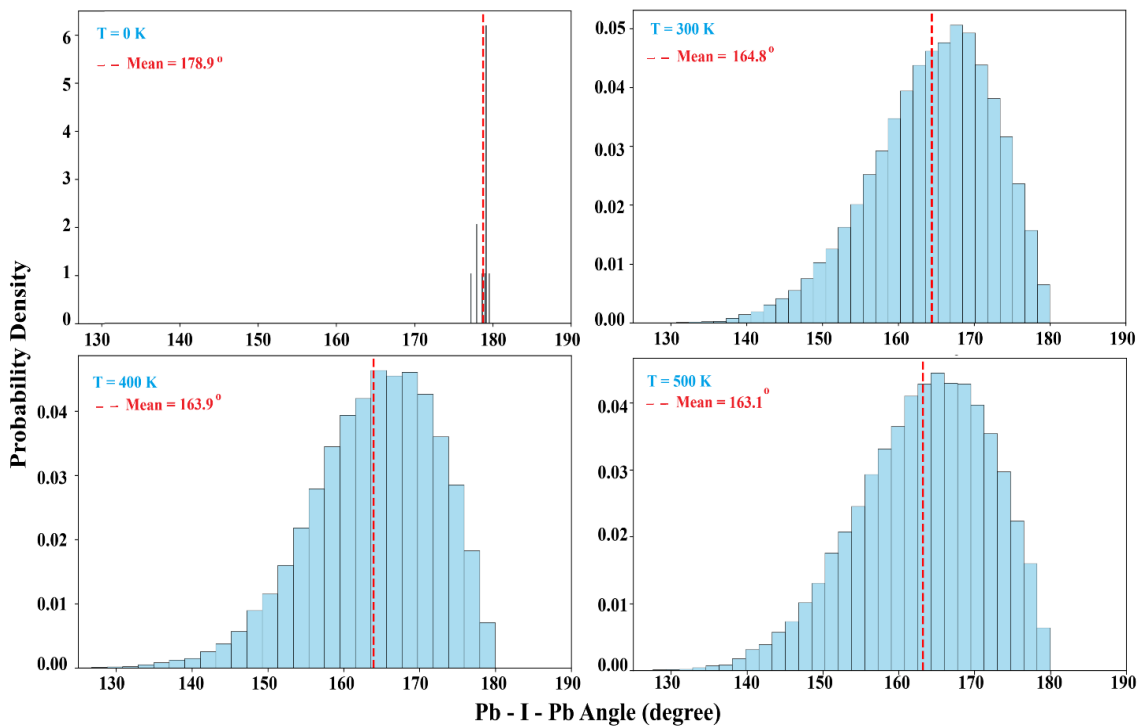

**Figure 5:** Evolution of Pb-I-Pb bond angle with temperature sampled over 20 ps followed by an initial 5 ps equilibration period.

## Conclusion:

In this study, density functional theory combined with ab initio molecular dynamics simulation (AIMD) was employed to study the distribution of structural motifs in the tetragonal phase of $FAPbI_3$. The study shows the microscopic origin in structural transformation that results in transforming and modulating the energy landscape of the electronic properties. Our results show that symmetry breaking induced by thermal excitation broadens the $PbI_6$ octahedral volume and the Pb–I–Pb bond angle due to enhanced FA cation orientation. These octahedral distortions give rise to a non-monotonic temperature induced band gap characteristic, with spectral blue shift at 300K, associated with diverse local environment due to symmetry breaking. The electronic structure investigation reveals that this dynamic structural phenomenon is directly responsible for modifying and modulating the electronic properties in $FAPbI_3$. Our findings highlight the role of local motifs in governing the electronic properties of hybrid perovskites, providing valuable insights for the rational design of more efficient materials.

## Acknowledgements

**H. Joshi and M. Wlazlo** acknowledges Polish high-performance computing infrastructure PLGrid (HPC Center: ACK Cyfronet AGH) for providing computer facilities and support within computational grant no. PLG/2024/017906.

## Conflict of Interest

The authors have no conflict to declare.

## Declaration of generative AI

The authors utilized ChatGPT to check grammar, spelling and writing style. The use of this AI tool was strictly limited to improving existing texts. No new content including texts or images was created through these tools.

## Author Contributions

**H. Joshi:** Data curation (lead); Formal analysis (equal); Investigation (lead); Methodology (lead); Validation (equal); Writing – original draft (lead). **K. C. Bhamu:** Data curation (supporting); Formal analysis (equal); Investigation (supporting); Methodology (supporting); Validation (supporting); Writing – original draft (supporting). **A. Shankar:** Data curation (supporting); Formal analysis (supporting); Investigation (supporting); Methodology (supporting); Validation (supporting); Writing – original draft (supporting). **Rana Biswas:** Data curation (supporting); Formal analysis (equal); Investigation (equal); Writing – original draft (supporting), Formal analysis (supporting); Investigation (equal); Methodology (supporting); Validation (equal), **M. Wlazło:** Data curation (equal); Formal analysis (lead); Investigation (lead); Methodology (equal); Validation (equal); Writing – original draft (supporting).